%% file: main.tex
\documentclass[a4paper,11pt]{article}
\usepackage{aaskaiid}
\usepackage{orcidlink}
\usepackage{tikz}
\usepackage{longtable}
\newcommand\subI{_{_\textsc{\tiny i}}}
\newcommand\subQ{_{_\textsc{\tiny q}}}
\newcommand\subU{_{_\textsc{\tiny u}}}
\newcommand\subV{_{_\textsc{\tiny v}}}
\renewcommand{\d}[1]{\ensuremath{\operatorname{d}\!{#1}}}

\title{Bridging Theory and Observation in the SKA Era: A Cosmological Polarized Radiative Transfer Framework for Point-to-Point Polarized Sky Comparisons}
\ShortTitle{CPRT: Bridging Theory and Observation in the SKA Era}

\author[1,2]{Jennifer Y. H. Chan\,\orcidlink{0000-0003-0314-7027}}
\ShortName{Chan et al.} 
\author[3, 4, 5]{Alvina Y. L. On\,\orcidlink{0000-0003-4479-4415}}
\author[5]{Paul C. W. Lai\,\orcidlink{0000-0003-3601-5127}}
\author[5]{Kinwah Wu\,
\orcidlink{0000-0002-7568-8765}}

\affiliation[1]{Department of Physics and Astronomy, Oberlin College, Oberlin, OH 44074, USA}
\affiliation[2]{Dunlap Institute for Astronomy and Astrophysics, University of Toronto, 50 St George St, Toronto, ON M5S 3H4, Canada}
\emailAdd{ychan1@oberlin.edu (JYHC)}
\affiliation[3]{Physics Division, National Center for Theoretical Sciences, Taipei 106319, Taiwan (ROC)}
\affiliation[4]{Center for Theory-Computation-Data Science Research, National Tsing Hua University, Hsinchu 30013, Taiwan (ROC)}
\affiliation[5]{Mullard Space Science Laboratory, University College London, Holmbury St. Mary, Surrey RH5 6NT, United Kingdom}
\emailAdd{alvina.on@ucl.ac.uk (AYLO)}
\emailAdd{chong.lai.22@ucl.ac.uk(PCWL)}
\emailAdd{kinwah.wu@ucl.ac.uk (KW)}

\abstract{
Realizing the full scientific potential of the SKA requires not only revolutionary instrumentation but also accurate modeling of light propagation in an evolving, expanding Universe, in order to translate intensity and polarization data into physical insight about magnetic fields and cosmic plasma. 
When all-sky cosmological polarized radiative transfer (CPRT) calculations meets SKA observations, theory and data
interlock to deliver a predictive, and testable picture of the evolving magneto-ionic Universe. This synergy transforms polarization observations---assembled into empirical maps of diffuse emission and rotation-measure (RM) grids of discrete sources---from descriptive data products into powerful astrophysical probes, advancing our understanding of cosmic magnetism across space and time.

The CPRT formalism---derived from fundamental conservation laws and incorporating relativistic, cosmological, and full radiative-transfer effects---provides a robust platform and a common framework for observers, theorists, and simulation experts to pursue shared scientific goals. Observers gain synthetic templates to interpret RM grids and polarization maps; theorists can directly confront models of magnetogenesis and magnetic-field evolution with data; and simulation experts obtain a post-processing tool to transform cosmological magneto-hydrodynamic (MHD) outputs into observable skies. Furthermore, CPRT serves as a powerful testbed when traditional RM-based methods reach their limitations---for example, in interpreting complex Faraday spectra, disentangling multiple intervening magnetized media, or achieving a coherent picture when diverse observational diagnostics---such as dispersion measure, synchrotron emission and spectral index, and dust polarization---are combined.
}

\begin{document}
\maketitle

\section{Introduction}

Magnetic fields are a fundamental but elusive component of the Universe.
They permeate every known cosmic environment, from stellar interiors and galactic disks to the tenuous filaments of the intergalactic medium that form the cosmic web \citep[see also][]{Sun01.2026.SKA, Mao01.2026.SKA, Curtin01.2026.SKA}. They are crucial in shaping the behavior of baryonic matter: guiding charged-particle transport, influencing gas dynamics and thermal balance, mediating feedback from stars and active galactic nuclei, and governing the acceleration and confinement of cosmic rays \citep{Beck2015AA, Subramanian2016RPPh}. On cosmological scales, magnetic fields are expected to be broadly coupled to baryonic flows, resulting in a spatial correlation with the filamentary cosmic web defined by the underlying dark-matter distribution (e.g. \citealt{Ryu2008Sci, Vazza2014MNRAS}), though this correspondence can be modified by turbulence, reconnection, and feedback processes.
Yet the origin, strength, and structure of these vast magnetic fields remain poorly constrained \citep{Vazza2018MNRAS, Bruggen2020IAUGA}. 
Understanding this hidden magnetic component is essential for a complete physical picture of structure formation, galaxy evolution, and cosmic plasma processes.

In parallel, advances in both observational capability and theoretical modeling have set the stage for a breakthrough in our understanding of cosmic magnetism, from compact sources to the largest structures in the Universe.

On the observational side, broadband polarimetric observations from the pathfinders and precursors to the Square Kilometre Array (SKA)---such as the Australian SKA Pathfinder \citep[ASKAP;][]{McConnell2020PASA, Gaensler2025PASA}, 
the Low-Frequency Array \citep[LOFAR;][]{vanHaarlem2013AA, Shimwell2017AA, Shimwell2026AA}
, the MeerKAT array \citep{Jonas2016mks, Taylor2024AJ}, 
and the Murchison Widefield Array \citep[MWA;][]{Tingay2013PASA, Lenc2016ApJ}---have enabled wide-field, high-sensitivity polarization studies across a broad frequency range. These observations, together with complementary measurements from other instruments, have revealed a wealth of polarized phenomena, ranging from Faraday-complex Galactic structures \citep[e.g.][]{Marijke2015ASSL_Bk_MW, 
Dickey2022ApJ, Heywood2022ApJ}, 
odd radio circles \citep[e.g.][]{Norris2025MNRAS, SabyasachiPal02.2026.SKA}, 
 to extreme range of magneto-ionic conditions traced by extragalactic fast radio bursts \citep[e.g.][]{Michilli2018Natur, Xu2022Natur, Pandhi2024ApJ, Pandhi2025ApJ}. 
These discoveries have demonstrated how wide-band, high-dynamic-range polarimetry can now probe magnetic fields across vastly different physical scales, cumulating to the SKA. 
They have also demonstrated that the magnetized Universe is far richer and more intricate than previously known, while highlighting the need for next-generation tools to interpret this growing wealth of data. 

On the theoretical front, cosmological magneto-hydrodynamic (MHD) simulations have matured into powerful tools for exploring the evolution of cosmic magnetic fields from early seeding to large-scale amplification. 
These simulations generate ``theoretical universes'' by evolving a cubic comoving volume through cosmological time, tracking the coupled dynamics of plasma, turbulence, and magnetic fields. State-of-the-art codes such as ENZO \citep{Bryan2014ApJS}, RAMSES \citep{Teyssier2002AA,Fromang2006AA}, and AREPO \citep{Springel2010MNRAS, Pakmor2013MNRAS, Pillepich2018MNRAS} solve the MHD equations self-consistently, revealing how turbulence, shear flows, and feedback processes strengthen and distribute magnetic energy throughout the cosmic web \citep{Vazza2014MNRAS, Marinacci2018MNRAS, Pakmor2024MNRAS}. With adaptive and moving-mesh techniques, these models bridge scales from megaparsec filaments to sub-kiloparsec galactic cores, permitting magnetic fields to emerge as dynamically active agents that influence gas accretion, feedback, and the thermal history of the Universe.

However, connecting such simulated volumes directly to real observations is non-trivial. 
The radiation detected by telescopes at a given observing frequency does not represent the sequential evolution of a single simulation cube; rather, it arises from light emitted by spatially distinct regions of the Universe at different lookback times. Observational data are therefore an integration along the observer’s past light cone---a convolution of emission from sources at varying emitted epochs. In contrast, simulation snapshots correspond to comoving volumes evaluated at discrete cosmic times, which cannot be simply stacked or cross-correlated to what an observer would measure along the light cone in polarized emission. Simple visualizations or synthetic map-making based on a snapshot of simulation cubes neglect this causal structure and risk introducing artificial artifacts. 

When emission from different epochs within the same comoving domain is cross-correlated, one effectively compares a region of the Universe with itself at multiple evolutionary stages. This self-mapping produces spurious coherence and artificial patterns, a limitation that can be understood in topological terms as akin to the constraints implied by Brouwer’s fixed-point theorem \citep[see e.g.][]{Starr11,Farmakis13}. These artifacts affect our interpretation of simulated observables and lead to an unjust comparison between simulations and data.

To overcome these limitations and establish a direct, physically consistent link between simulations and polarization observations, a covariant Cosmological Polarized Radiative Transfer (CPRT) formalism \citep{Chan2019MNRAS, On2019MNRAS} has been developed. It constructs observables directly on the observer’s past light cone, tracing the full Stokes parameters 
$(I_\nu,Q_\nu,U_\nu,V_\nu)$ along sightlines through an evolving, magnetized universe. By explicitly incorporating all relevant radiative processes---emission, absorption, Faraday rotation, and Faraday conversion (a propagation effect that converts linear polarization into circular polarization, and vice versa)---and by self-consistently accounting for both local and global effects in a fully covariant, frequency-dependent manner, CPRT ensures that each line of sight captures radiation from the correct cosmic epoch and spatial location, thereby bringing theoretical simulation frontiers onto the observational plane.

Observationally, polarization data are typically interpreted {using Rotation Measure (RM) synthesis \citep{Brentjens2005AA, VanEck2026ApJS, Carcamo01.2026.SKA} and $QU$-fitting \citep[e.g.][]{OSullivan2012MNRAS, Oberhelman2026MNRAS}. RM synthesis reconstructs the Faraday depth distribution from the observed polarization data, whereas $QU$ fitting constrains parameterized models of the magneto-ionic medium through direct fitting of the Stokes spectra.} In the simplest case---where a single, Faraday rotating screen lies in front of a polarized background source---the full CPRT formalism reduces to the classical linear relation between polarization angle and wavelength squared, from which the RM is defined. In this sense, RM can be understood as a special case of CPRT under highly simplified physical conditions.

More generally, the observed complex polarization, $P(\lambda^2)$ ($= Q(\lambda^2)+iU(\lambda^2)$) can be expressed in terms of the Faraday dispersion function, $F(\phi)$, which describes the distribution of polarized emission as a function of Faraday depth $\phi$ along the line of sight. This relation is derived within a forward model in which the observed polarization arises from the linear superposition of emission components undergoing Faraday rotation. However, real astrophysical environments often violate these simplifying assumptions, 
particularly when synchrotron emission and Faraday rotation are co-spatial within inhomogeneous magneto-ionic media, giving rise to Faraday-thick structures and complex, wavelength-dependent depolarization. While multiple Faraday-thin components can often be treated as a linear superposition, such mixed emitting and rotating regions introduce additional ambiguities. {Under these circumstances, reconstructing or interpreting the underlying magneto-ionic structure becomes increasingly non-unique. RM synthesis may suffer from ambiguities in Faraday depth space, while QU fitting can exhibit degeneracies between competing parametric models. The CPRT framework provides the complementary, physically self-consistent radiative-transfer approach:} starting from physical models of magnetized plasma, it predicts the full frequency-dependent Stokes observables that telescopes measure.

The CPRT formalism generates full-Stokes observables and synthetic RMs from magnetized media, matched to telescope's frequency coverage. When forward-modeled through SKA’s measurement equations---the mathematical description of how the telescope records the sky---CPRT makes it possible to compare simulated and observed data on equal terms. Specifically, it allows 
\begin{itemize}
    \item Direct comparison of synthetic and observed polarized sky, or RM maps,
    \item Deep insights and understanding of depolarization and polarization enhancement behaviors,
    \item Quantitative analysis of the relative importance of different physical mechanisms at play in different environments and cosmological epochs, 
    \item Testing of how well the RM grid recovers true underlying magnetic-field statistics \citep[see also][]{OSullivan01.2026.SKA, Loi01.2026.SKA},
    \item Generating high-fidelity polarization templates on which advanced statistical and inverse-modeling techniques can be applied to reconstruct the three-dimensional magnetic morphology and its evolutionary history, thereby providing insights to discriminate between primordial and astrophysical seeding scenarios.
\end{itemize}

 This approach provides a realistic and observation-ready framework that connects theory and measurement, integrating cosmological MHD simulation advances, realistic modeling of observables, and the broadband polarimetric capabilities of the SKA. In doing so, it establishes a physically grounded basis for interpreting SKA polarization data and RM synthesis and $QU$-fitting results. {By linking physically motivated forward modeling with observational inference}, CPRT closes the loop between simulations and observations, enabling robust predictions and self-consistent interpretation of the magneto-ionic Universe across cosmic time and scales.

\section{The CPRT Formalism, Rotation Measure, and Cosmic Magnetism Diagnostics}

At the heart of the CPRT framework lies the fundamental equation governing how polarized radiation is generated, absorbed, and modified as it propagates through magnetized plasma in an expanding Universe. The CPRT equation, in the co-moving frame, reads: 
\begin{equation}
\frac{\rm d}{\d z}
\begin{bmatrix} 
{\mathcal I}\\
{\mathcal Q}\\
{\mathcal U}\\
{\mathcal V}
\end{bmatrix} 
=   
 ({1+z})
 \left\{
 -    \begin{bmatrix} \kappa & q & u & v \\ q &\kappa & f & -g \\ u & -f & \kappa&h  \\ v& g & -h& \kappa  \end{bmatrix} 
\begin{bmatrix} 
{\mathcal I}\\
{\mathcal Q}\\
{\mathcal U}\\
{\mathcal V}
\end{bmatrix} 
 +\begin{bmatrix}
 \epsilon\subI \\ \epsilon\subQ \\ \epsilon\subU \\ \epsilon\subV
 \end{bmatrix} 
\frac{1}{\nu^3} \right\} 
\frac{\d s}{\d z}    
\label{eq:covarPRTinz}
\end{equation} 
\citep{Chan2019MNRAS}, where 
    $\kappa,\, q,\, u,\, v$ are the absorption coefficients,
    $\epsilon$ are the emission coefficients,  
    $f$ is the Faraday rotation coefficient, 
    $g$ and $h$ are the Faraday conversion coefficients. 
The invariant Stokes parameters are related to the usual Stokes parameters by 
$[\,{\mathcal I}\, {\mathcal Q}\, {\mathcal U}\, {\mathcal V}\,]^{\rm T}  = [\,I\, Q\, U\, V\,]^{\rm T} /\nu^{3}$\,.
The radiation path length $s$ changes with respect to redshift $z$ in a flat Friedmann-Robertson-Walker (FRW) universe is given by 
\begin{equation} 
\frac{\d s}{\d z} 
= 
\frac{c}{H_{0}}\,({1+z})^{-1}
   \left[\Omega_{{\rm r},0}(1+z)^{4}+\Omega_{{\rm m},0}(1+z)^{3}+\Omega_{\Lambda,0}\right]^{-\frac{1}{2}}  
\label{eq:dldz}
\end{equation}
\citep[see e.g.\,][]{Peacock1999}, 
where $H_{0}$ is the Hubble parameter, $\Omega_{{\rm r},0}$, $\Omega_{{\rm m},0}$ and $\Omega_{\Lambda,0}$ are the dimensionless energy densities of relativistic matter and radiation, non-relativistic matter, 
and a cosmological constant (dark energy with an equation of state of $w \equiv -1$), respectively. 
The subscript ``0" denotes that the quantities are measured at present\,(i.e. $z=0$).  

The widely used magnetic diagnostics, Rotation Measure (RM) ${\cal R}$, is defined as 
\begin{equation} 
{\cal R}  =  (\Delta \varphi) \;\! \lambda^{-2} =   \left(\varphi - \varphi_0\right)\;\! \lambda^{-2} \ , 
\end{equation}    
where $\Delta \varphi$ is the rotation of the linear polarization angle, and $\lambda$ the observing radiation wavelength. 
RM is stemmed from Equation~\ref{eq:covarPRTinz}. The full derivation can be found in \citet{On2019MNRAS}. Here we summarize the key steps to highlight the different assumptions used to derive the RM quantity. 

If cosmological effects are negligible, Equation~\ref{eq:covarPRTinz} reduced to the standard PRT equation: 
\begin{equation} 
  \frac{\rm d}{{\rm d} s}   
   \begin{bmatrix} I \\ Q \\ U \\ V \end{bmatrix} 
   =  
    -    \begin{bmatrix} \kappa & q & u & v \\ q &\kappa & f & -g \\ u & -f & \kappa&h  \\ v& g & -h& \kappa  \end{bmatrix} 
         \begin{bmatrix} I \\ Q \\ U \\ V \end{bmatrix} 
        +   \begin{bmatrix}
        \epsilon\subI \\ \epsilon\subQ \\ \epsilon\subU \\ \epsilon\subV 
        \end{bmatrix}   \  , 
\label{eq-PRT}
\end{equation}

When emission and absorption are negligible and Faraday conversion can be ignored, the simplified two-component $(Q,U)$ transfer reduces to
\begin{equation} 
  \frac{\rm d}{{\rm d} s}   \begin{bmatrix}  Q \\ U  \end{bmatrix}   =  
    -    \begin{bmatrix}    0 & f \\   -f & 0   \end{bmatrix} 
         \begin{bmatrix}  Q \\ U  \end{bmatrix}    \  .   
\label{eq-PRT_restrictive}
\end{equation}  
It has the sole Faraday rotation coefficient $f$, which is determined by the properties of free electrons and the magnetic field along the line-of-sight. 
For plasma consisting of both thermal electrons and non-thermal electrons, 
  the Faraday rotation coefficient is given by 
\begin{align}
f_{\rm tot} = f_{\rm th} + f_{\rm nt} 
  = \left(  \frac{\omega_{\rm p}^2 \cos \theta}{ c\, \omega_{\rm B}} 
    \left(\frac{\omega_{\rm B}^2}{\omega^2}  \right)   \right)   + 
    \left(
     \frac{1}{\uppi } \left(\frac{e^3}{m_{\rm e}^2c^4} \right)  \zeta (p, \gamma_{\rm i}) \, 
     n_{\rm e,nt} \, {B}_\parallel \,  \lambda^2 
\right) 
\end{align} 
  \citep{Pacholczyk1977, Jones1977a}, in the limit of $\omega^2 \gg \omega_{\rm B}^2$. Here, 
   $n_{\rm e,th}$ is the thermal electron number density,
   $n_{\rm e,nt}$ is the non-thermal electron number density, 
   $B_\parallel = \cos\theta$ is the magnetic field strength along the line of sight,  where $\theta$ is the angle between the field and the line of sight; 
  $\omega = 2 \uppi \nu$ is the angular frequency of radiation,
  $\omega_{\rm p} = (4\uppi n_{\rm e,th} e^2/m_{\rm e})^{1/2}$ is the plasma frequency, 
  $\omega_{\rm B} = (e B/ m_{\rm e}c)$ is the electron gyro-frequency, 
  the factor $
   \zeta (p, \gamma_{\rm i})  =  \frac{(p-1)(p+2)}{(p+1)}  \left( \frac{\ln \gamma_{\rm i}}{{\gamma_{\rm i}}^2}  \right)$, for $p >1$, 
    assuming an isotropic distribution of non-thermal electrons with a power-law energy spectrum of index $p$, with $\gamma_{\rm i}$ the low-energy cut-off.  
   Using Equation~\ref{eq-PRT_restrictive} and the definition of linear polarization angle, it can be shown that the change in the linear polarization angle along the line-of-sight is 
\begin{equation} 
  \frac{{\rm d}\varphi}{{\rm d} s}  =  \frac{1}{2}\left(\frac{1}{U^2+Q^2}\right)
     \left(Q \frac{{\rm d}U}{{\rm d} s}-U\frac{{\rm d}Q}{{\rm d} s} \right)   = \frac{f}{2} \ . 
\label{eq-dphids}
\end{equation} 
In a sufficiently weak magnetic field where $\omega_{\rm B} \ll \omega$, 
if the plasma has only thermal electrons, or non-thermal electron number density is insignificant, 
  a direct integration of equation~({\ref{eq-dphids}}) with $f = f_{\rm th}$ yields 
\begin{equation} 
  \varphi(s)    =    \varphi_0 + \frac{2\uppi e^3}{m_{\rm e}^2(c\, \omega)^2} 
     \int^s_{s_0} {\rm d}s' \, n_{\rm e, th}(s') \, B_\parallel(s') \ . 
\end{equation} 
If non-thermal electron number density is non-negligible, the RM is given by 
\begin{equation} 
 {\cal R}(s) =  \frac{e^3}{2\uppi m_{\rm e}^2c^4}   \int^s_{s_0} {\rm d}s' \ n_{\rm e}(s') \, \Theta (s') \, {B}_\parallel(s')  \ , 
\label{eq-R_x}
\end{equation}  
  where
  $n_{\rm e}$ is the total electron number density, and 
  the weighting factor of different populations of $n_{\rm e}$ contributing to the Faraday rotation effect being 
    $\Theta(s) = 1 - \Upsilon(s) \, [1- \zeta(p, \gamma_{\rm i})]\big\vert_{s}$, with $\Upsilon(s)$ the local fraction of non-thermal electrons. If only thermal electrons are present, $\Upsilon(s)=0$ such that $\Theta(s) = 1$,
   the widely-used formula in RM analysis
   of magnetized astrophysical media is recovered: 
\begin{equation} 
 {\cal R}(s)  =   0.812  \int^s_{s_0} \frac{{\rm d}s'}{{\rm pc}} \!
      \left(\frac{n_{\rm e, th}(s')}{{\rm cm}^{-3}} \right)\! \left( \frac{{B}_\parallel(s')}{\mu{\rm G}}\right) \,  {\rm rad}\;\!{\rm m}^{-2} \  . 
\label{eq:RM_int}
\end{equation} 

As such, RM---or the Faraday depth obtained through Rotation Measure Synthesis--- is not only a derived observable constructed from Stokes $Q$, and $U$, but its interpretation in terms of physical magneto-ionic conditions along the line of sight implicitly assumes that all the simplifying conditions listed above are valid. 
RM synthesis reconstructs the Faraday depth spectrum by Fourier transforming the complex polarization as a function of $\lambda^2$, providing a powerful diagnostic of multi-component structures. Complementarily, $QU$-fitting models the observed $Q(\lambda^2)$ and $U(\lambda^2)$ spectra using parametric physical or phenomenological descriptions \citep[e.g.][]{OSullivan2012MNRAS}. Both approaches relax the single-screen assumption and are useful tools for identifying Faraday complexity, such as multiple emitting or rotating regions or co-spatial emission and rotation. 
However, to achieve full, unambiguous interpretive power, theoretical modeling of the complete Stokes spectra and polarization maps is needed to disentangle the contributions from different components and propagation effects. In this context, the CPRT framework provides a fundamentally different approach: rather than inferring structure from observables, it forward-models the generation and propagation of polarized radiation by incorporating emission, absorption, Faraday rotation, and Faraday conversion self-consistently within an evolving cosmological setting.

\begin{figure}
  \centering
  \begin{tikzpicture}[scale=0.8, node distance = 0.25, auto]
    \coordinate (1rayNode) at (135:2.9cm);
    \coordinate (2rayNode) at (45:2.4cm);
    \coordinate (3rayNode) at (225:2.4cm);
    \coordinate (4rayNode) at (315:2.4cm);
    \coordinate (originNode) at (0:0cm);

 \draw[<-, >=stealth, very thick] (1rayNode.south) -- (0,0) node[left,pos=-.05]{$z$} 
 node[left,pos=0.2]{}  
 node[left,pos=0.4]{} 
 node[left,pos=0.6]{} 
 node[left,pos=0.8]{} 
 node[left,pos=0.95]{$z=0$};
 \draw[-> , >=stealth, purple, very thick] (2rayNode.south) -- (0,0) 
  node[ purple, right, pos=-0.1]{{$(z, \theta, \phi)$}}
 node[ purple, left,pos=0]{} 
 node[left,pos=0.2]{} 
 node[ purple, left,pos=0.4]{} 
 node[ purple, left,pos=0.6]{} node[left,pos=0.8]{};

    \draw[fill,color=purple] (barycentric cs:2rayNode=1.0,originNode=0) circle (2pt);
    \draw[fill,color=purple] (barycentric cs:2rayNode=0.75,originNode=0.25) circle (2pt);
    \draw[fill,color=purple] (barycentric cs:2rayNode=0.5,originNode=0.5) circle (2pt);
    \draw[fill,color=purple] (barycentric cs:2rayNode=0.25,originNode=0.75) circle (2pt);
    \draw[fill,color=purple] (barycentric cs:2rayNode=0,originNode=1.0) circle (2pt);

    \draw[black!50] (0,0) circle (2.4cm);
    \draw[black!50] (0,0) circle (1.8cm);
    \draw[black!50] (0,0) circle (1.2cm);
    \draw[black!50] (0,0) circle (0.6cm);

  \end{tikzpicture}
  \caption[Illustration of the ray-tracing concept of the all-sky algorithm]{An illustration of an all-sky CPRT framework employing a ray-tracing method. The CPRT equation is solved along each light ray (shown in red), which is labelled by three parameters: the cosmological redshift $z$, which links to the light travelled distance (or time) and the position on the celestial sky $(\theta, \phi)$. The observer is located at the center ($z=0$). .} 
  \label{fig:algo}
\end{figure}
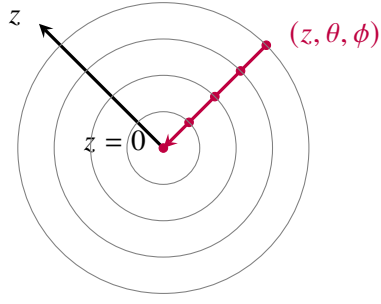

The CPRT framework makes this possible by allowing different types of ray-tracing calculations: from single-ray configurations (well suited to distant, compact sources), to pencil-beam calculations---i.e., a narrow bundle of neighboring rays sampling a small solid angle---for extended astrophysical objects and diffuse media, and to full all-sky implementations (see Fig.~\ref{fig:algo}) for studies of the Milky Way, and the cosmic web. 

\begin{figure}[h]
    \centering
	\includegraphics[width=1.00\columnwidth]{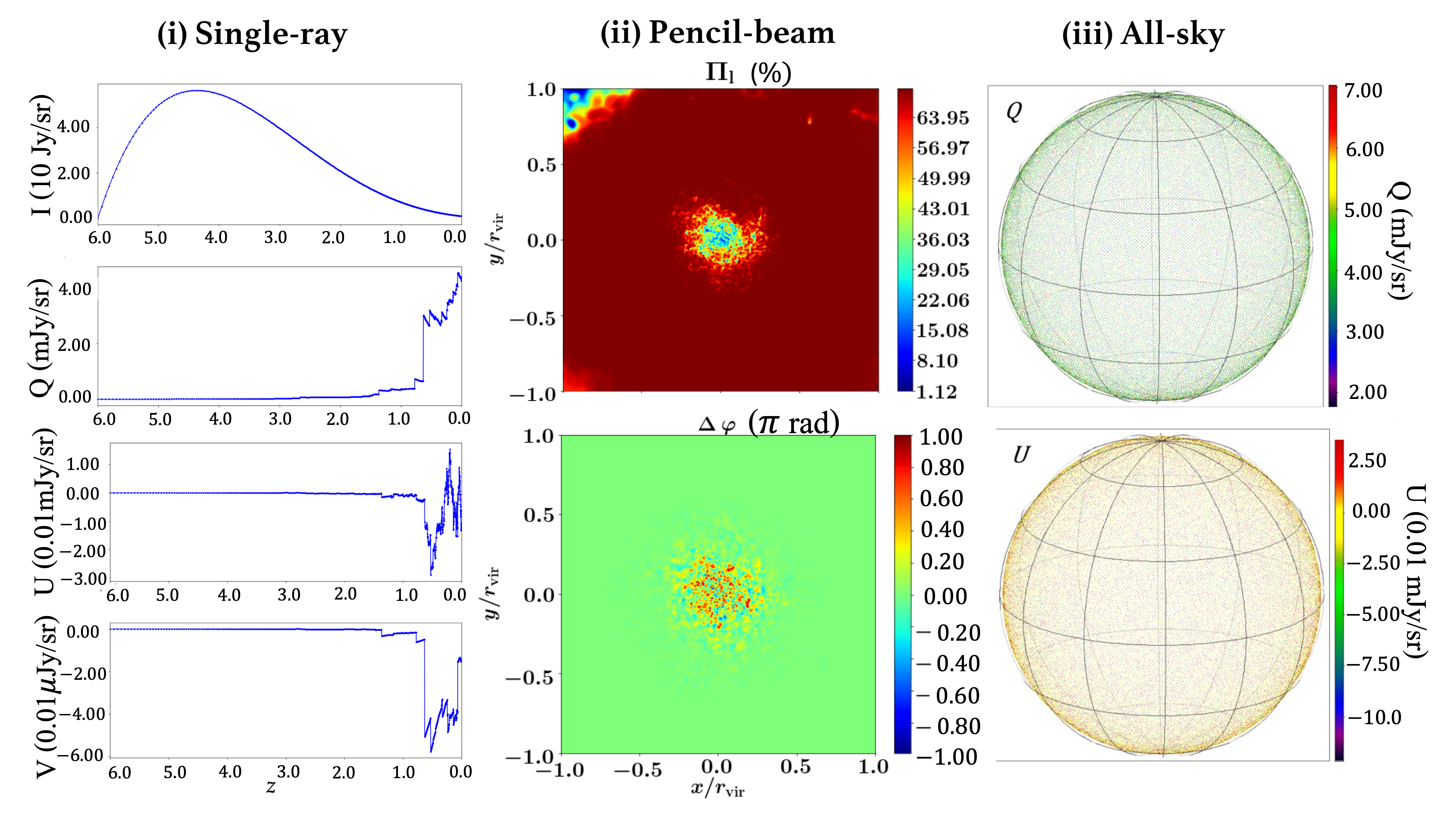}
    \caption{Examples of CPRT data products for various astrophysical studies. Column (i) shows the redshift evolution of $I$ (intensity), $Q$ and $U$ (linear polarization), and $V$ (circular polarization) along a ray propagating through highly rarefied intergalactic space. The rapid variations in polarization at $z\lesssim 0.8$ arise from Faraday rotation and conversion as the ray encounters increasingly magnetized and denser structures at lower redshift. Column (ii) presents pencil-beam maps of the degree of linear polarization and the change in polarization angle for a simulated galaxy cluster, showing spatial variations driven by the underlying magneto-ionic structures. Column (iii) displays the all-sky linear polarization predicted by a magnetic universe model, demonstrating CPRT’s ability to connect local plasma processes with global observable signatures. A higher-resolution version of these figures and further details of the underlying model can be found in \citet{Chan2019MNRAS}.}
    \label{fig:CPRT_DataProduct}
\end{figure}

Examples of some of the CPRT data products for different astrophysical applications are illustrated in Fig.~\ref{fig:CPRT_DataProduct}. Column (i) illustrates the redshift evolution of the Stokes parameters along a single ray propagating through highly rarefied intergalactic plasma, where the rapid variations in $Q, U, V$ at low redshift arise as the ray encounters denser and more strongly magnetized regions. Column (ii) presents pencil-beam maps of the degree of linear polarization and the change in polarization angle for a simulated galaxy cluster
    \citep{Barnes2018MNRAS}, 
demonstrating how spatial variations in the magneto-ionic medium translate into observable polarization structure. Column (iii) shows an all-sky linear polarization obtained by applying the CPRT formalism to a magnetic universe model \citep{Chan2019MNRAS}, in which cosmological MHD simulation outputs are mapped onto the observer’s past light cone to construct a full-sky polarized signal.

This flexibility enables CPRT to bridge the gap between detailed theoretical modeling and broadband polarimetric observations across all cosmic environments, 
and to provide physically grounded predictions for interpreting or validating results from RM synthesis and $QU$-fitting, particularly in regimes where spectral and spatial polarization complexity leads to degeneracies or non-unique solutions.

\section{Scientific Gains and Synergies Between Observations, Simulations, and Theory}

\textbf{The Cosmological Polarized Radiative Transfer (CPRT) framework} represents a major step toward uniting theoretical modeling and observation in the SKA era. By generating synthetic all-sky full Stokes spectra, polarization maps, and RM grids, CPRT enables true apple-to-apple comparison between simulations and observations--- allowing pixel-by-pixel, statistically robust tests of magneto-ionic models. This shared framework places observers and theorists within the same interpretive space, ensuring that both communities can assess magnetic-field predictions on equal observational footing.

\textbf{Constraining the origins and evolution of cosmic magnetic fields} will be one of the key frontiers enabled by this approach \citep[see also][]{OSullivan01.2026.SKA, Vacca02.2026.SKA}. By linking cosmological MHD simulations to polarization observables, CPRT enables different magnetic-field seeding scenarios---such as primordial and astrophysical origins---to be tested and constrained through their predicted statistical signatures in the distribution of polarized sources, polarization fractions, and Faraday-depth structures observed with the SKA. However, the differences between primordial and astrophysical seeding are often subtle and can overlap, particularly in regions dominated by nonlinear evolution such as galaxy clusters. The strongest discriminants are expected to emerge from large-scale, low-density environments--- filaments and voids ---and from the observed redshift evolution of magnetic-field strength and coherence.

\textbf{Beyond standard RM methods}, CPRT moves past the assumptions of a single (or a series of) Faraday-rotating screen(s) and a simple polarized background. By incorporating emission, absorption, Faraday rotation, and conversion in a unified framework, it can handle complex, multi-component sightlines and multi-phase media in an expanding Universe. This allows the validity of inverse-modeling methods, such as traditional RM synthesis and $QU$-fitting, to be quantitatively assessed under realistic conditions.

\textbf{The SKA’s dense RM grid} will provide millions of polarized background sources that act as backlights shining through intervening magneto-ionic material. At the same time, \textbf{diffuse polarized synchrotron emission} from the Milky Way \citep[see e.g.,][]{Ma01.2026.SKA, Sun01.2026.SKA}, galaxy clusters \citep[see e.g.,][]{Cassano01.2026.SKA, Loi01.2026.SKA, Vacca01.2026.SKA}, and the cosmic web \citep[see e.g.][]{Cuciti01.2026.SKA, Dabhade01.2026.SKA, Tabatabaei01.2026.SKA} forms the polarized fabric of the sky. By combining these complementary observables---the discrete backlights and the continuous diffuse emissions---and anchoring them with CPRT predictions, a robust assessment of, and pathway toward the reconstruction of large-scale magnetized structures becomes possible. 

\textbf{Foreground separation} is a critical step in this process. Joint modeling of Galactic and extragalactic magnetized plasma within the CPRT framework provides self-consistent physical templates for disentangling overlapping foregrounds, accurately modeling frequency-dependent depolarization and polarization enhancement, quantifying biases, and isolating the extragalactic magnetic signal with minimal contamination.

Furthermore, \textbf{Faraday tomography and multi-component analysis} \citep[see also][]{Carcamo01.2026.SKA, Tahani01.2026.SKA} will benefit from the SKA’s broad $\lambda^2$ coverage and extended baselines, which together enhance angular and Faraday-depth resolution, while its large collecting area and wide bandwidth improve sensitivity---enabling reconstruction of complex Faraday-depth spectra and the disentangling of overlapping magnetized regions along the line of sight. CPRT naturally outputs the full Stokes evolution, capturing the $Q(\lambda^2)$ and $U(\lambda^2)$ dependencies of cosmologically evolving media---beyond the simplifying assumptions of Burn’s slab model \citep{Burn1966MNRAS} and its variants---and explicitly accounting for the sequential ordering of magnetized regions and their cumulative effects on the observed polarization. This capability offers physically grounded testbeds to assess the reconstruction accuracy of multi-component plasmas, depolarization effects, and Faraday-thick emission, thereby calibrating ambiguities and establishing benchmarks for SKA analysis pipelines, while also informing the optimization of SKA survey strategies.

Last but not the least, because CPRT naturally includes cosmic expansion and evolving plasma conditions, it allows polarization data \textbf{across redshift} to be compared with theoretical predictions for magnetic-field growth, enabling \textbf{tests of amplification histories and redshift evolution}.

\section{Summary}
Looking ahead, advancing cosmic magnetism requires that instrumentation, simulations, modeling, and theory progress shoulder to shoulder to fully realize the scientific potential of the SKA and its pathfinders. The CPRT framework provides a robust and precise platform for this collective endeavor—linking theoretical models of cosmologically evolving plasmas with synthetic observations and instrumental responses in a self-consistent and unambiguous manner. Its capability to interface comoving simulation cubes and compute directly for the polarized sky represents a major breakthrough, enabling straightforward and quantitative comparison between theory and observation. 

Within this framework, classical RM analysis emerges as a simplified, Faraday-thin limit, while CPRT serves as the forward-modeling and validation toolset that complements {observational inference techniques} such as RM synthesis and $QU$-fitting. In practice, RM-based methods remain effective for simple sightlines. However, CPRT becomes essential in the broadband, multi-component, and cosmological regimes probed by the SKA---where Faraday complexity, line-of-sight mixing, and light-cone effects cannot be neglected. In these cases, CPRT provides the physically consistent basis required to interpret polarization spectra, test magneto-ionic models, and quantify reconstruction uncertainties.

By placing forward modeling and inverse inference within a unified physical context, CPRT closes the loop between simulations and observations. This enables robust, testable predictions and supports precision studies of astrophysical and cosmological magnetism, ensuring that the SKA’s transformative capabilities translate into physically grounded constraints on the origin and evolution of cosmic magnetic fields. 

\bibliographystyle{abbrvnat-maxbibnames4}
\bibliography{chapter}

\end{document}

%% file: chapter.bib
@incollection{Carcamo01.2026.SKA, author = {Miguel Carcamo and author2 and author3 and author4 and author5},title = {},year = {2026},publisher = {},note = {arXiv search: Report number AASKAII/Carcamo01},booktitle = {Advancing Astrophysics with the SKA -- II (AASKAII)}}

@incollection{Ma01.2026.SKA, author = {Yik Ki Ma and author2 and author3 and author4 and author5},title = {},year = {2026},publisher = {},note = {arXiv search: Report number AASKAII/Ma01},booktitle = {Advancing Astrophysics with the SKA -- II (AASKAII)}}

@incollection{Sun01.2026.SKA, author = {Xiaohui Sun and author2 and author3 and author4 and author5},title = {},year = {2026},publisher = {},note = {arXiv search: Report number AASKAII/Sun01},booktitle = {Advancing Astrophysics with the SKA -- II (AASKAII)}}

@incollection{Mao01.2026.SKA, author = {Sui Ann Mao and author2 and author3 and author4 and author5},title = {},year = {2026},publisher = {},note = {arXiv search: Report number AASKAII/Mao01},booktitle = {Advancing Astrophysics with the SKA -- II (AASKAII)}}

@incollection{Tahani01.2026.SKA, author = {Mehrnoosh Tahani and author2 and author3 and author4 and author5},title = {},year = {2026},publisher = {},note = {arXiv search: Report number AASKAII/Tahani01},booktitle = {Advancing Astrophysics with the SKA -- II (AASKAII)}}

@incollection{Loi01.2026.SKA, author = {Francesca Loi and author2 and author3 and author4 and author5},title = {},year = {2026},publisher = {},note = {arXiv search: Report number AASKAII/Loi01},booktitle = {Advancing Astrophysics with the SKA -- II (AASKAII)}}

@incollection{Cassano01.2026.SKA, author = {Rossella Cassano and author2 and author3 and author4 and author5},title = {},year = {2026},publisher = {},note = {arXiv search: Report number AASKAII/Cassano01},booktitle = {Advancing Astrophysics with the SKA -- II (AASKAII)}}

@incollection{Cuciti01.2026.SKA, author = {Virginia Cuciti and author2 and author3 and author4 and author5},title = {},year = {2026},publisher = {},note = {arXiv search: Report number AASKAII/Cuciti01},booktitle = {Advancing Astrophysics with the SKA -- II (AASKAII)}}

@incollection{Dabhade01.2026.SKA, author = {Pratik Dabhade and author2 and author3 and author4 and author5},title = {},year = {2026},publisher = {},note = {arXiv search: Report number AASKAII/Dabhade01},booktitle = {Advancing Astrophysics with the SKA -- II (AASKAII)}}

@incollection{Tabatabaei01.2026.SKA, author = {Fatemeh S. Tabatabaei and author2 and author3 and author4 and author5},title = {},year = {2026},publisher = {},note = {arXiv search: Report number AASKAII/Tabatabaei01},booktitle = {Advancing Astrophysics with the SKA -- II (AASKAII)}}

@incollection{OSullivan01.2026.SKA, author = {Shane P. O'Sullivan and author2 and author3 and author4 and author5},title = {},year = {2026},publisher = {},note = {arXiv search: Report number AASKAII/OSullivan01},booktitle = {Advancing Astrophysics with the SKA -- II (AASKAII)}}

@incollection{Vacca01.2026.SKA, author = {Valentina Vacca and author2 and author3 and author4 and author5},title = {},year = {2026},publisher = {},note = {arXiv search: Report number AASKAII/Vacca01},booktitle = {Advancing Astrophysics with the SKA -- II (AASKAII)}}

@incollection{Vacca02.2026.SKA, author = {Valentina Vacca and author2 and author3 and author4 and author5},title = {},year = {2026},publisher = {},note = {arXiv search: Report number AASKAII/Vacca02},booktitle = {Advancing Astrophysics with the SKA -- II (AASKAII)}}

@incollection{SabyasachiPal02.2026.SKA, author = {Sabyasachi Pal and author2 and author3 and author4 and author5},title = {},year = {2026},publisher = {},note = {arXiv search: Report number AASKAII/SabyasachiPal02},booktitle = {Advancing Astrophysics with the SKA -- II (AASKAII)}}

@incollection{Curtin01.2026.SKA, author = {Alice P. Curtin and author2 and author3 and author4 and author5},title = {},year = {2026},publisher = {},note = {arXiv search: Report number AASKAII/Curtin01},booktitle = {Advancing Astrophysics with the SKA -- II (AASKAII)}}

@article{Barnes2018MNRAS,
	adsurl = {https://ui.adsabs.harvard.edu/abs/2018MNRAS.476.2890B},
	archiveprefix = {arXiv},
	author = {{Barnes}, David J. and {On}, Alvina Y.~L. and {Wu}, Kinwah and {Kawata}, Daisuke},
	doi = {10.1093/mnras/sty400},
	eprint = {1811.08861},
	journal = {\mnras},
	month = may,
	number = {3},
	pages = {2890-2904},
	primaryclass = {astro-ph.CO},
	title = {{SPMHD simulations of structure formation}},
	volume = {476},
	year = 2018}

@ARTICLE{Burn1966MNRAS,
       author = {{Burn}, B.~J.},
        title = "{On the depolarization of discrete radio sources by Faraday dispersion}",
      journal = {\mnras},
         year = 1966,
        month = jan,
       volume = {133},
        pages = {67},
          doi = {10.1093/mnras/133.1.67},
       adsurl = {https://ui.adsabs.harvard.edu/abs/1966MNRAS.133...67B}
}

@ARTICLE{Chan2019MNRAS,
       author = {{Chan}, Jennifer Y.~H. and {Wu}, Kinwah and {On}, Alvina Y.~L. and {Barnes}, David J. and {McEwen}, Jason D. and {Kitching}, Thomas D.},
        title = "{Covariant polarized radiative transfer on cosmological scales for investigating large-scale magnetic field structures}",
      journal = {\mnras},
         year = 2019,
        month = apr,
       volume = {484},
       number = {2},
        pages = {1427-1455},
          doi = {10.1093/mnras/sty3498},
archivePrefix = {arXiv},
       eprint = {1901.04581},
 primaryClass = {astro-ph.CO},
       adsurl = {https://ui.adsabs.harvard.edu/abs/2019MNRAS.484.1427C}
}

@ARTICLE{On2019MNRAS,
       author = {{On}, Alvina Y.~L. and {Chan}, Jennifer Y.~H. and {Wu}, Kinwah and {Saxton}, Curtis J. and {van Driel-Gesztelyi}, Lidia},
        title = "{Polarized radiative transfer, rotation measure fluctuations, and large-scale magnetic fields}",
      journal = {\mnras},
         year = 2019,
        month = dec,
       volume = {490},
       number = {2},
        pages = {1697-1713},
          doi = {10.1093/mnras/stz2683},
archivePrefix = {arXiv},
       eprint = {1909.06703},
 primaryClass = {astro-ph.HE},
       adsurl = {https://ui.adsabs.harvard.edu/abs/2019MNRAS.490.1697O}
}

@BOOK{Pacholczyk1977,
   author = {{Pacholczyk}, A.~G.},
    title = "{Radio galaxies: Radiation transfer, dynamics, stability and evolution of a synchrotron plasmon}",
    publisher = {Pergamon Press, Oxford},
     year = 1977,
   adsurl = {http://ukads.nottingham.ac.uk/abs/1977OISNP..89.....P}
}

@ARTICLE{Jones1977a,
   author = {{Jones}, T.~W. and {O'Dell}, S.~L.},
    title = "{Transfer of polarized radiation in self-absorbed synchrotron sources. I. Results for a homogeneous source}",
  journal = {\apj},
     year = 1977,
    month = jun,
   volume = 214,
    pages = {522},
      doi = {10.1086/155278},
   adsurl = {http://ukads.nottingham.ac.uk/abs/1977ApJ...214..522J}
}

@BOOK{Peacock1999,
   author = {{Peacock}, J.~A.},
    title = "{Cosmological Physics}",
booktitle = {Cosmological Physics, by John A.~Peacock, pp.~704.~ISBN 052141072X.~Cambridge, UK: Cambridge University Press, January 1999.},
     year = 1999,
    month = jan,
    pages = {704},
    isbn = {9780521422703},
    lccn = {98029460},
    series = {Cambridge Astrophysics},
    publisher = {Cambridge Univ. Press},
   adsurl = {http://adsabs.harvard.edu/abs/1999coph.book.....P}
}

@inbook{Starr11, 
booktitle={General equilibrium theory: An introduction},
edition={Second}, 
title={The Brouwer Fixed-Point Theorem}, DOI={10.1017/CBO9780511975356.014}, 
publisher={Cambridge University Press, Cambridge}, 
location ={Cambridge}, 
author={Starr, R.~M.}, 
year={2011}, 
pages={99–108}
}

@book{Farmakis13,
       author = {Farmakis, I. and {Moskowitz}, M.~A.},
        title = "{Fixed point theorems and their applications}",
     publisher = "World Scientific Publishing, Singapore",
     location = "Singapore",
     year = "2013"
}

@ARTICLE{Vazza2014MNRAS,
       author = {{Vazza}, F. and {Br{\"u}ggen}, M. and {Gheller}, C. and {Wang}, P.},
        title = "{On the amplification of magnetic fields in cosmic filaments and galaxy clusters}",
      journal = {\mnras},
         year = 2014,
        month = dec,
       volume = {445},
       number = {4},
        pages = {3706-3722},
          doi = {10.1093/mnras/stu1896},
archivePrefix = {arXiv},
       eprint = {1409.2640},
 primaryClass = {astro-ph.CO},
       adsurl = {https://ui.adsabs.harvard.edu/abs/2014MNRAS.445.3706V}
}

@ARTICLE{Marinacci2018MNRAS,
       author = {{Marinacci}, Federico and {Vogelsberger}, Mark and {Pakmor}, R{\"u}diger and {Torrey}, Paul and {Springel}, Volker and {Hernquist}, Lars and {Nelson}, Dylan and {Weinberger}, Rainer and {Pillepich}, Annalisa and {Naiman}, Jill and {Genel}, Shy},
        title = "{First results from the IllustrisTNG simulations: radio haloes and magnetic fields}",
      journal = {\mnras},
         year = 2018,
        month = nov,
       volume = {480},
       number = {4},
        pages = {5113-5139},
          doi = {10.1093/mnras/sty2206},
archivePrefix = {arXiv},
       eprint = {1707.03396},
 primaryClass = {astro-ph.CO},
       adsurl = {https://ui.adsabs.harvard.edu/abs/2018MNRAS.480.5113M}
}

@INPROCEEDINGS{Bruggen2020IAUGA,
       author = {{Br{\"u}ggen}, Marcus and {O'Sullivan}, Shane and {Bonafede}, Annalisa and {Vazza}, Franco},
        title = "{Magnetic fields in the intergalactic medium and in the cosmic web}",
    booktitle = {IAU General Assembly},
         year = 2020,
        month = mar,
        pages = {303-306},
          doi = {10.1017/S1743921319004460},
       adsurl = {https://ui.adsabs.harvard.edu/abs/2020IAUGA..30..303B}
}

@ARTICLE{Vazza2018MNRAS,
       author = {{Vazza}, F. and {Brunetti}, G. and {Br{\"u}ggen}, M. and {Bonafede}, A.},
        title = "{Resolved magnetic dynamo action in the simulated intracluster medium}",
      journal = {\mnras},
         year = 2018,
        month = feb,
       volume = {474},
       number = {2},
        pages = {1672-1687},
          doi = {10.1093/mnras/stx2830},
archivePrefix = {arXiv},
       eprint = {1711.02673},
 primaryClass = {astro-ph.CO},
       adsurl = {https://ui.adsabs.harvard.edu/abs/2018MNRAS.474.1672V}
}

@ARTICLE{Beck2015AA,
       author = {{Beck}, Rainer},
        title = "{Magnetic fields in the nearby spiral galaxy IC 342: A multi-frequency radio polarization study}",
      journal = {\aap},
         year = 2015,
        month = jun,
       volume = {578},
          eid = {A93},
        pages = {A93},
          doi = {10.1051/0004-6361/201425572},
archivePrefix = {arXiv},
       eprint = {1502.05439},
 primaryClass = {astro-ph.GA},
       adsurl = {https://ui.adsabs.harvard.edu/abs/2015A&A...578A..93B}
}

@ARTICLE{Subramanian2016RPPh,
       author = {{Subramanian}, Kandaswamy},
        title = "{The origin, evolution and signatures of primordial magnetic fields}",
      journal = {Reports on Progress in Physics},
         year = 2016,
        month = jul,
       volume = {79},
       number = {7},
          eid = {076901},
        pages = {076901},
          doi = {10.1088/0034-4885/79/7/076901},
archivePrefix = {arXiv},
       eprint = {1504.02311},
 primaryClass = {astro-ph.CO},
       adsurl = {https://ui.adsabs.harvard.edu/abs/2016RPPh...79g6901S}
}

@ARTICLE{Michilli2018Natur,
       author = {{Michilli}, D. and {Seymour}, A. and {Hessels}, J.~W.~T. and {Spitler}, L.~G. and {Gajjar}, V. and {Archibald}, A.~M. and {Bower}, G.~C. and {Chatterjee}, S. and {Cordes}, J.~M. and {Gourdji}, K. and {Heald}, G.~H. and {Kaspi}, V.~M. and {Law}, C.~J. and {Sobey}, C. and {Adams}, E.~A.~K. and {Bassa}, C.~G. and {Bogdanov}, S. and {Brinkman}, C. and {Demorest}, P. and {Fernandez}, F. and {Hellbourg}, G. and {Lazio}, T.~J.~W. and {Lynch}, R.~S. and {Maddox}, N. and {Marcote}, B. and {McLaughlin}, M.~A. and {Paragi}, Z. and {Ransom}, S.~M. and {Scholz}, P. and {Siemion}, A.~P.~V. and {Tendulkar}, S.~P. and {van Rooy}, P. and {Wharton}, R.~S. and {Whitlow}, D.},
        title = "{An extreme magneto-ionic environment associated with the fast radio burst source FRB 121102}",
      journal = {\nat},
         year = 2018,
        month = jan,
       volume = {553},
       number = {7687},
        pages = {182-185},
          doi = {10.1038/nature25149},
archivePrefix = {arXiv},
       eprint = {1801.03965},
 primaryClass = {astro-ph.HE},
       adsurl = {https://ui.adsabs.harvard.edu/abs/2018Natur.553..182M}
}

@ARTICLE{Xu2022Natur,
       author = {{Xu}, H. and {Niu}, J.~R. and {Chen}, P. and {Lee}, K.~J. and {Zhu}, W.~W. and {Dong}, S. and {Zhang}, B. and {Jiang}, J.~C. and {Wang}, B.~J. and {Xu}, J.~W. and {Zhang}, C.~F. and {Fu}, H. and {Filippenko}, A.~V. and {Peng}, E.~W. and {Zhou}, D.~J. and {Zhang}, Y.~K. and {Wang}, P. and {Feng}, Y. and {Li}, Y. and {Brink}, T.~G. and {Li}, D.~Z. and {Lu}, W. and {Yang}, Y.~P. and {Caballero}, R.~N. and {Cai}, C. and {Chen}, M.~Z. and {Dai}, Z.~G. and {Djorgovski}, S.~G. and {Esamdin}, A. and {Gan}, H.~Q. and {Guhathakurta}, P. and {Han}, J.~L. and {Hao}, L.~F. and {Huang}, Y.~X. and {Jiang}, P. and {Li}, C.~K. and {Li}, D. and {Li}, H. and {Li}, X.~Q. and {Li}, Z.~X. and {Liu}, Z.~Y. and {Luo}, R. and {Men}, Y.~P. and {Niu}, C.~H. and {Peng}, W.~X. and {Qian}, L. and {Song}, L.~M. and {Stern}, D. and {Stockton}, A. and {Sun}, J.~H. and {Wang}, F.~Y. and {Wang}, M. and {Wang}, N. and {Wang}, W.~Y. and {Wu}, X.~F. and {Xiao}, S. and {Xiong}, S.~L. and {Xu}, Y.~H. and {Xu}, R.~X. and {Yang}, J. and {Yang}, X. and {Yao}, R. and {Yi}, Q.~B. and {Yue}, Y.~L. and {Yu}, D.~J. and {Yu}, W.~F. and {Yuan}, J.~P. and {Zhang}, B.~B. and {Zhang}, S.~B. and {Zhang}, S.~N. and {Zhao}, Y. and {Zheng}, W.~K. and {Zhu}, Y. and {Zou}, J.~H.},
        title = "{A fast radio burst source at a complex magnetized site in a barred galaxy}",
      journal = {\nat},
         year = 2022,
        month = sep,
       volume = {609},
       number = {7928},
        pages = {685-688},
          doi = {10.1038/s41586-022-05071-8},
archivePrefix = {arXiv},
       eprint = {2111.11764},
 primaryClass = {astro-ph.HE},
       adsurl = {https://ui.adsabs.harvard.edu/abs/2022Natur.609..685X}
}

@ARTICLE{Pandhi2024ApJ,
       author = {{Pandhi}, Ayush and {Pleunis}, Ziggy and {Mckinven}, Ryan and {Gaensler}, B.~M. and {Su}, Jianing and {Ng}, Cherry and {Bhardwaj}, Mohit and {Brar}, Charanjot and {Cassanelli}, Tomas and {Cook}, Amanda and {Curtin}, Alice P. and {Kaspi}, Victoria M. and {Lazda}, Mattias and {Leung}, Calvin and {Li}, Dongzi and {Masui}, Kiyoshi W. and {Michilli}, Daniele and {Nimmo}, Kenzie and {Pearlman}, Aaron B. and {Petroff}, Emily and {Rafiei-Ravandi}, Masoud and {Sand}, Ketan R. and {Scholz}, Paul and {Shin}, Kaitlyn and {Smith}, Kendrick and {Stairs}, Ingrid},
        title = "{Polarization Properties of 128 Nonrepeating Fast Radio Bursts from the First CHIME/FRB Baseband Catalog}",
      journal = {\apj},
         year = 2024,
        month = jun,
       volume = {968},
       number = {2},
          eid = {50},
        pages = {50},
          doi = {10.3847/1538-4357/ad40aa},
archivePrefix = {arXiv},
       eprint = {2401.17378},
 primaryClass = {astro-ph.HE},
       adsurl = {https://ui.adsabs.harvard.edu/abs/2024ApJ...968...50P}
}

@ARTICLE{Heywood2022ApJ,
       author = {{Heywood}, I. and {Rammala}, I. and {Camilo}, F. and {Cotton}, W.~D. and {Yusef-Zadeh}, F. and {Abbott}, T.~D. and {Adam}, R.~M. and {Adams}, G. and {Aldera}, M.~A. and {Asad}, K.~M.~B. and {Bauermeister}, E.~F. and {Bennett}, T.~G.~H. and {Bester}, H.~L. and {Bode}, W.~A. and {Botha}, D.~H. and {Botha}, A.~G. and {Brederode}, L.~R.~S. and {Buchner}, S. and {Burger}, J.~P. and {Cheetham}, T. and {de Villiers}, D.~I.~L. and {Dikgale-Mahlakoana}, M.~A. and {du Toit}, L.~J. and {Esterhuyse}, S.~W.~P. and {Fanaroff}, B.~L. and {February}, S. and {Fourie}, D.~J. and {Frank}, B.~S. and {Gamatham}, R.~R.~G. and {Geyer}, M. and {Goedhart}, S. and {Gouws}, M. and {Gumede}, S.~C. and {Hlakola}, M.~J. and {Hokwana}, A. and {Hoosen}, S.~W. and {Horrell}, J.~M.~G. and {Hugo}, B. and {Isaacson}, A.~I. and {J{\'o}zsa}, G.~I.~G. and {Jonas}, J.~L. and {Joubert}, A.~F. and {Julie}, R.~P.~M. and {Kapp}, F.~B. and {Kenyon}, J.~S. and {Kotz{\'e}}, P.~P.~A. and {Kriek}, N. and {Kriel}, H. and {Krishnan}, V.~K. and {Lehmensiek}, R. and {Liebenberg}, D. and {Lord}, R.~T. and {Lunsky}, B.~M. and {Madisa}, K. and {Magnus}, L.~G. and {Mahgoub}, O. and {Makhaba}, A. and {Makhathini}, S. and {Malan}, J.~A. and {Manley}, J.~R. and {Marais}, S.~J. and {Martens}, A. and {Mauch}, T. and {Merry}, B.~C. and {Millenaar}, R.~P. and {Mnyandu}, N. and {Mokone}, O.~J. and {Monama}, T.~E. and {Mphego}, M.~C. and {New}, W.~S. and {Ngcebetsha}, B. and {Ngoasheng}, K.~J. and {Ockards}, M.~T. and {Oozeer}, N. and {Otto}, A.~J. and {Passmoor}, S.~S. and {Patel}, A.~A. and {Peens-Hough}, A. and {Perkins}, S.~J. and {Ramaila}, A.~J.~T. and {Ramanujam}, N.~M.~R. and {Ramudzuli}, Z.~R. and {Ratcliffe}, S.~M. and {Robyntjies}, A. and {Salie}, S. and {Sambu}, N. and {Schollar}, C.~T.~G. and {Schwardt}, L.~C. and {Schwartz}, R.~L. and {Serylak}, M. and {Siebrits}, R. and {Sirothia}, S.~K. and {Slabber}, M. and {Smirnov}, O.~M. and {Sofeya}, L. and {Taljaard}, B. and {Tasse}, C. and {Tiplady}, A.~J. and {Toruvanda}, O. and {Twum}, S.~N. and {van Balla}, T.~J. and {van der Byl}, A. and {van der Merwe}, C. and {Van Tonder}, V. and {Van Wyk}, R. and {Venter}, A.~J. and {Venter}, M. and {Wallace}, B.~H. and {Welz}, M.~G. and {Williams}, L.~P. and {Xaia}, B.},
        title = "{The 1.28 GHz MeerKAT Galactic Center Mosaic}",
      journal = {\apj},
         year = 2022,
        month = feb,
       volume = {925},
       number = {2},
          eid = {165},
        pages = {165},
          doi = {10.3847/1538-4357/ac449a},
archivePrefix = {arXiv},
       eprint = {2201.10541},
 primaryClass = {astro-ph.GA},
       adsurl = {https://ui.adsabs.harvard.edu/abs/2022ApJ...925..165H}
}

@ARTICLE{Brentjens2005AA,
       author = {{Brentjens}, M.~A. and {de Bruyn}, A.~G.},
        title = "{Faraday rotation measure synthesis}",
      journal = {\aap},
         year = 2005,
        month = oct,
       volume = {441},
       number = {3},
        pages = {1217-1228},
          doi = {10.1051/0004-6361:20052990},
archivePrefix = {arXiv},
       eprint = {astro-ph/0507349},
 primaryClass = {astro-ph},
       adsurl = {https://ui.adsabs.harvard.edu/abs/2005A&A...441.1217B}
}

@ARTICLE{McConnell2020PASA,
       author = {{McConnell}, D. and {Hale}, C.~L. and {Lenc}, E. and {Banfield}, J.~K. and {Heald}, George and {Hotan}, A.~W. and {Leung}, James K. and {Moss}, Vanessa A. and {Murphy}, Tara and {O'Brien}, Andrew and {Pritchard}, Joshua and {Raja}, Wasim and {Sadler}, Elaine M. and {Stewart}, Adam and {Thomson}, Alec J.~M. and {Whiting}, M. and {Allison}, James R. and {Amy}, S.~W. and {Anderson}, C. and {Ball}, Lewis and {Bannister}, Keith W. and {Bell}, Martin and {Bock}, Douglas C.-J. and {Bolton}, Russ and {Bunton}, J.~D. and {Chippendale}, A.~P. and {Collier}, J.~D. and {Cooray}, F.~R. and {Cornwell}, T.~J. and {Diamond}, P.~J. and {Edwards}, P.~G. and {Gupta}, N. and {Hayman}, Douglas B. and {Heywood}, Ian and {Jackson}, C.~A. and {Koribalski}, B{\"a}rbel S. and {Lee-Waddell}, Karen and {McClure-Griffiths}, N.~M. and {Ng}, Alan and {Norris}, Ray P. and {Phillips}, Chris and {Reynolds}, John E. and {Roxby}, Daniel N. and {Schinckel}, Antony E.~T. and {Shields}, Matt and {Tremblay}, Chenoa and {Tzioumis}, A. and {Voronkov}, M.~A. and {Westmeier}, Tobias},
        title = "{The Rapid ASKAP Continuum Survey I: Design and first results}",
      journal = {\pasa},
         year = 2020,
        month = nov,
       volume = {37},
          eid = {e048},
        pages = {e048},
          doi = {10.1017/pasa.2020.41},
archivePrefix = {arXiv},
       eprint = {2012.00747},
 primaryClass = {astro-ph.IM},
       adsurl = {https://ui.adsabs.harvard.edu/abs/2020PASA...37...48M}
}

@ARTICLE{vanHaarlem2013AA,
       author = {{van Haarlem}, M.~P. and {Wise}, M.~W. and {Gunst}, A.~W. and {Heald}, G. and {McKean}, J.~P. and {Hessels}, J.~W.~T. and {de Bruyn}, A.~G. and {Nijboer}, R. and {Swinbank}, J. and {Fallows}, R. and {Brentjens}, M. and {Nelles}, A. and {Beck}, R. and {Falcke}, H. and {Fender}, R. and {H{\"o}randel}, J. and {Koopmans}, L.~V.~E. and {Mann}, G. and {Miley}, G. and {R{\"o}ttgering}, H. and {Stappers}, B.~W. and {Wijers}, R.~A.~M.~J. and {Zaroubi}, S. and {van den Akker}, M. and {Alexov}, A. and {Anderson}, J. and {Anderson}, K. and {van Ardenne}, A. and {Arts}, M. and {Asgekar}, A. and {Avruch}, I.~M. and {Batejat}, F. and {B{\"a}hren}, L. and {Bell}, M.~E. and {Bell}, M.~R. and {van Bemmel}, I. and {Bennema}, P. and {Bentum}, M.~J. and {Bernardi}, G. and {Best}, P. and {B{\^\i}rzan}, L. and {Bonafede}, A. and {Boonstra}, A.-J. and {Braun}, R. and {Bregman}, J. and {Breitling}, F. and {van de Brink}, R.~H. and {Broderick}, J. and {Broekema}, P.~C. and {Brouw}, W.~N. and {Br{\"u}ggen}, M. and {Butcher}, H.~R. and {van Cappellen}, W. and {Ciardi}, B. and {Coenen}, T. and {Conway}, J. and {Coolen}, A. and {Corstanje}, A. and {Damstra}, S. and {Davies}, O. and {Deller}, A.~T. and {Dettmar}, R.-J. and {van Diepen}, G. and {Dijkstra}, K. and {Donker}, P. and {Doorduin}, A. and {Dromer}, J. and {Drost}, M. and {van Duin}, A. and {Eisl{\"o}ffel}, J. and {van Enst}, J. and {Ferrari}, C. and {Frieswijk}, W. and {Gankema}, H. and {Garrett}, M.~A. and {de Gasperin}, F. and {Gerbers}, M. and {de Geus}, E. and {Grie{\ss}meier}, J.-M. and {Grit}, T. and {Gruppen}, P. and {Hamaker}, J.~P. and {Hassall}, T. and {Hoeft}, M. and {Holties}, H.~A. and {Horneffer}, A. and {van der Horst}, A. and {van Houwelingen}, A. and {Huijgen}, A. and {Iacobelli}, M. and {Intema}, H. and {Jackson}, N. and {Jelic}, V. and {de Jong}, A. and {Juette}, E. and {Kant}, D. and {Karastergiou}, A. and {Koers}, A. and {Kollen}, H. and {Kondratiev}, V.~I. and {Kooistra}, E. and {Koopman}, Y. and {Koster}, A. and {Kuniyoshi}, M. and {Kramer}, M. and {Kuper}, G. and {Lambropoulos}, P. and {Law}, C. and {van Leeuwen}, J. and {Lemaitre}, J. and {Loose}, M. and {Maat}, P. and {Macario}, G. and {Markoff}, S. and {Masters}, J. and {McFadden}, R.~A. and {McKay-Bukowski}, D. and {Meijering}, H. and {Meulman}, H. and {Mevius}, M. and {Middelberg}, E. and {Millenaar}, R. and {Miller-Jones}, J.~C.~A. and {Mohan}, R.~N. and {Mol}, J.~D. and {Morawietz}, J. and {Morganti}, R. and {Mulcahy}, D.~D. and {Mulder}, E. and {Munk}, H. and {Nieuwenhuis}, L. and {van Nieuwpoort}, R. and {Noordam}, J.~E. and {Norden}, M. and {Noutsos}, A. and {Offringa}, A.~R. and {Olofsson}, H. and {Omar}, A. and {Orr{\'u}}, E. and {Overeem}, R. and {Paas}, H. and {Pandey-Pommier}, M. and {Pandey}, V.~N. and {Pizzo}, R. and {Polatidis}, A. and {Rafferty}, D. and {Rawlings}, S. and {Reich}, W. and {de Reijer}, J.-P. and {Reitsma}, J. and {Renting}, G.~A. and {Riemers}, P. and {Rol}, E. and {Romein}, J.~W. and {Roosjen}, J. and {Ruiter}, M. and {Scaife}, A. and {van der Schaaf}, K. and {Scheers}, B. and {Schellart}, P. and {Schoenmakers}, A. and {Schoonderbeek}, G. and {Serylak}, M. and {Shulevski}, A. and {Sluman}, J. and {Smirnov}, O. and {Sobey}, C. and {Spreeuw}, H. and {Steinmetz}, M. and {Sterks}, C.~G.~M. and {Stiepel}, H.-J. and {Stuurwold}, K. and {Tagger}, M. and {Tang}, Y. and {Tasse}, C. and {Thomas}, I. and {Thoudam}, S. and {Toribio}, M.~C. and {van der Tol}, B. and {Usov}, O. and {van Veelen}, M. and {van der Veen}, A.-J. and {ter Veen}, S. and {Verbiest}, J.~P.~W. and {Vermeulen}, R. and {Vermaas}, N. and {Vocks}, C. and {Vogt}, C. and {de Vos}, M. and {van der Wal}, E. and {van Weeren}, R. and {Weggemans}, H. and {Weltevrede}, P. and {White}, S. and {Wijnholds}, S.~J. and {Wilhelmsson}, T. and {Wucknitz}, O. and {Yatawatta}, S. and {Zarka}, P. and {Zensus}, A.},
        title = "{LOFAR: The LOw-Frequency ARray}",
      journal = {\aap},
         year = 2013,
        month = aug,
       volume = {556},
          eid = {A2},
        pages = {A2},
          doi = {10.1051/0004-6361/201220873},
archivePrefix = {arXiv},
       eprint = {1305.3550},
 primaryClass = {astro-ph.IM},
       adsurl = {https://ui.adsabs.harvard.edu/abs/2013A&A...556A...2V}
}

@ARTICLE{Shimwell2017AA,
       author = {{Shimwell}, T.~W. and {R{\"o}ttgering}, H.~J.~A. and {Best}, P.~N. and {Williams}, W.~L. and {Dijkema}, T.~J. and {de Gasperin}, F. and {Hardcastle}, M.~J. and {Heald}, G.~H. and {Hoang}, D.~N. and {Horneffer}, A. and {Intema}, H. and {Mahony}, E.~K. and {Mandal}, S. and {Mechev}, A.~P. and {Morabito}, L. and {Oonk}, J.~B.~R. and {Rafferty}, D. and {Retana-Montenegro}, E. and {Sabater}, J. and {Tasse}, C. and {van Weeren}, R.~J. and {Br{\"u}ggen}, M. and {Brunetti}, G. and {Chy{\.z}y}, K.~T. and {Conway}, J.~E. and {Haverkorn}, M. and {Jackson}, N. and {Jarvis}, M.~J. and {McKean}, J.~P. and {Miley}, G.~K. and {Morganti}, R. and {White}, G.~J. and {Wise}, M.~W. and {van Bemmel}, I.~M. and {Beck}, R. and {Brienza}, M. and {Bonafede}, A. and {Calistro Rivera}, G. and {Cassano}, R. and {Clarke}, A.~O. and {Cseh}, D. and {Deller}, A. and {Drabent}, A. and {van Driel}, W. and {Engels}, D. and {Falcke}, H. and {Ferrari}, C. and {Fr{\"o}hlich}, S. and {Garrett}, M.~A. and {Harwood}, J.~J. and {Heesen}, V. and {Hoeft}, M. and {Horellou}, C. and {Israel}, F.~P. and {Kapi{\'n}ska}, A.~D. and {Kunert-Bajraszewska}, M. and {McKay}, D.~J. and {Mohan}, N.~R. and {Orr{\'u}}, E. and {Pizzo}, R.~F. and {Prandoni}, I. and {Schwarz}, D.~J. and {Shulevski}, A. and {Sipior}, M. and {Smith}, D.~J.~B. and {Sridhar}, S.~S. and {Steinmetz}, M. and {Stroe}, A. and {Varenius}, E. and {van der Werf}, P.~P. and {Zensus}, J.~A. and {Zwart}, J.~T.~L.},
        title = "{The LOFAR Two-metre Sky Survey. I. Survey description and preliminary data release}",
      journal = {\aap},
         year = 2017,
        month = feb,
       volume = {598},
          eid = {A104},
        pages = {A104},
          doi = {10.1051/0004-6361/201629313},
archivePrefix = {arXiv},
       eprint = {1611.02700},
 primaryClass = {astro-ph.IM},
       adsurl = {https://ui.adsabs.harvard.edu/abs/2017A&A...598A.104S}
}

@ARTICLE{Shimwell2026AA,
       author = {{Shimwell}, T.~W. and {Hardcastle}, M.~J. and {Tasse}, C. and {Drabent}, A. and {Botteon}, A. and {Williams}, W.~L. and {Best}, P.~N. and {R{\"o}ttgering}, H.~J.~A. and {Br{\"u}ggen}, M. and {Brunetti}, G. and {Callingham}, J.~R. and {Chy{\.z}y}, K.~T. and {Conway}, J.~E. and {De Gasperin}, F. and {Haverkorn}, M. and {Horellou}, C. and {Jackson}, N. and {Miley}, G.~K. and {Morabito}, L.~K. and {Morganti}, R. and {O'Sullivan}, S.~P. and {Schwarz}, D.~J. and {Smith}, D.~J.~B. and {van Weeren}, R.~J. and {Vedantham}, H.~K. and {White}, G.~J. and {Ahmadi}, A. and {Alegre}, L. and {Arias}, M. and {Asabere}, B. and {Bahr-Kalus}, B. and {Barkus}, B. and {Bilicki}, M. and {B{\"o}hme}, L. and {Brentjens}, M. and {Brienza}, M. and {Bomans}, D.~J. and {Bonafede}, A. and {Bonato}, M. and {Bonnassieux}, E. and {Boxelaar}, J.~M. and {Camera}, S. and {Cassano}, R. and {Chilufya}, J. and {Cianfaglione}, M. and {Croston}, J.~H. and {Cuciti}, V. and {Dabhade}, P. and {De Rubeis}, E. and {de Jong}, J.~M.~G.~H.~J. and {Dallacasa}, D. and {Dettmar}, R.~J. and {Duncan}, K.~J. and {Di Gennaro}, G. and {Edler}, H.~W. and {Groeneveld}, C. and {G{\"u}rkan}, G. and {Hajduk}, M. and {Hale}, C.~L. and {Heesen}, V. and {Hoang}, D.~N. and {Hoeft}, M. and {Holties}, H. and {Horton}, M.~A. and {Iacobelli}, M. and {Jamrozy}, M. and {Jarvis}, M.~J. and {Jelic}, V. and {Kadler}, M. and {Kondapally}, R. and {Kunert-Bajraszewska}, M. and {Loose}, M. and {Magliocchetti}, M. and {Ma{\l}ek}, K. and {Manzano}, C. and {McKean}, J.~P. and {Mevius}, M. and {Mingo}, B. and {Miskolczi}, A. and {Misra}, A. and {Mold{\'o}n}, J. and {Nair}, D.~G. and {Nakoneczny}, S.~J. and {Orru}, E. and {Pashapour-Ahmadabadi}, M. and {Pasini}, T. and {Petley}, J. and {Pierce}, J.~C.~S. and {Prandoni}, I. and {Rafferty}, D. and {Rajpurohit}, K. and {Riseley}, C.~J. and {Roberts}, I.~D. and {Sethi}, S. and {Shulevski}, A. and {Stein}, M. and {Stuardi}, C. and {Sweijen}, F. and {ter Veen}, S. and {Timmerman}, R. and {Vaccari}, M. and {Wijnholds}, S.},
        title = "{The LOFAR Two-metre Sky Survey: VII. Third Data Release}",
      journal = {\aap},
         year = 2026,
        month = mar,
       volume = {707},
          eid = {A198},
        pages = {A198},
          doi = {10.1051/0004-6361/202557749},
archivePrefix = {arXiv},
       eprint = {2602.15949},
 primaryClass = {astro-ph.GA},
       adsurl = {https://ui.adsabs.harvard.edu/abs/2026A&A...707A.198S}
}

@ARTICLE{Norris2025MNRAS,
       author = {{Norris}, Ray P. and {Koribalski}, B{\"a}rbel S. and {Hale}, Catherine L. and {Jarvis}, Matt J. and {Macgregor}, Peter J. and {Taylor}, A. Russell},
        title = "{MeerKAT discovery of a MIGHTEE Odd Radio Circle}",
      journal = {\mnras},
         year = 2025,
        month = feb,
       volume = {537},
       number = {1},
        pages = {L42-L48},
          doi = {10.1093/mnrasl/slae114},
archivePrefix = {arXiv},
       eprint = {2411.17311},
 primaryClass = {astro-ph.GA},
       adsurl = {https://ui.adsabs.harvard.edu/abs/2025MNRAS.537L..42N}
}

@ARTICLE{Gaensler2025PASA,
       author = {{Gaensler}, B.~M. and {Heald}, G.~H. and {McClure-Griffiths}, N.~M. and {Anderson}, C.~S. and {Van Eck}, C.~L. and {West}, J.~L. and {Thomson}, A.~J.~M. and {Leahy}, J.~P. and {Rudnick}, L. and {Ma}, Y.~K. and {Akahori}, Takuya and {G{\"u}rkan}, G. and {Landecker}, T.~L. and {Mao}, S.~A. and {O'Sullivan}, S.~P. and {Raja}, W. and {Sun}, X. and {Vernstrom}, T. and {Baidoo}, Lerato and {Carretti}, Ettore and {Taylor}, A.~R. and {Willis}, A.~G. and {Osinga}, Erik and {Livingston}, J.~D. and {Alexander}, E.~L. and {Alonso-L{\'o}pez}, David and {Amaral}, A.~D. and {An}, T. and {Bracco}, Andrea and {Bradbury}, S. and {Br{\"u}ggen}, Marcus and {Eswaraiah}, Chakali and {En{\ss}lin}, Torsten and {Galvin}, T.~J. and {Haverkorn}, Marijke and {Hopkins}, A.~M. and {Hutschenreuter}, Sebastian and {Ideguchi}, Shinsuke and {Jaswanth}, S. and {Jung}, S. Lyla and {Kaczmarek}, J.~F. and {Kothes}, Roland and {Lazarevi{\'c}}, Sanja and {Leahy}, Denis and {Loi}, Francesca and {Marvil}, Joshua R. and {Norris}, Ray and {Pandhi}, Ayush and {Price}, Jason M. and {Riseley}, C.~J. and {Ryder}, P. and {Seta}, Amit and {Shaw}, Vasundhara and {Shen}, A.~X. and {Sobey}, C. and {Stil}, J. and {Stuardi}, Chiara and {Upasana}, Gupta and {Vanderwoude}, Shannon and {Velovi{\'c}}, Velibor},
        title = "{The Polarisation Sky Survey of the Universe's Magnetism (POSSUM): Science goals and survey description}",
      journal = {\pasa},
         year = 2025,
        month = jun,
       volume = {42},
          eid = {e091},
        pages = {e091},
          doi = {10.1017/pasa.2025.10031},
archivePrefix = {arXiv},
       eprint = {2505.08272},
 primaryClass = {astro-ph.GA},
       adsurl = {https://ui.adsabs.harvard.edu/abs/2025PASA...42...91G}
}

@ARTICLE{Taylor2024AJ,
       author = {{Taylor}, A.~R. and {Legodi}, L.~S.},
        title = "{A MeerKAT Polarization Survey of Southern Calibration Sources}",
      journal = {\aj},
         year = 2024,
        month = jun,
       volume = {167},
       number = {6},
          eid = {273},
        pages = {273},
          doi = {10.3847/1538-3881/ad4150},
archivePrefix = {arXiv},
       eprint = {2405.04131},
 primaryClass = {astro-ph.GA},
       adsurl = {https://ui.adsabs.harvard.edu/abs/2024AJ....167..273T}
}

@INPROCEEDINGS{Jonas2016mks,
       author = {{Jonas}, J. and {MeerKAT Team}},
        title = "{The MeerKAT Radio Telescope}",
    booktitle = {MeerKAT Science: On the Pathway to the SKA},
         year = 2016,
        month = jan,
          eid = {1},
        pages = {1},
          doi = {10.22323/1.277.0001},
       adsurl = {https://ui.adsabs.harvard.edu/abs/2016mks..confE...1J}
}

@ARTICLE{Tingay2013PASA,
       author = {{Tingay}, S.~J. and {Goeke}, R. and {Bowman}, J.~D. and {Emrich}, D. and {Ord}, S.~M. and {Mitchell}, D.~A. and {Morales}, M.~F. and {Booler}, T. and {Crosse}, B. and {Wayth}, R.~B. and {Lonsdale}, C.~J. and {Tremblay}, S. and {Pallot}, D. and {Colegate}, T. and {Wicenec}, A. and {Kudryavtseva}, N. and {Arcus}, W. and {Barnes}, D. and {Bernardi}, G. and {Briggs}, F. and {Burns}, S. and {Bunton}, J.~D. and {Cappallo}, R.~J. and {Corey}, B.~E. and {Deshpande}, A. and {Desouza}, L. and {Gaensler}, B.~M. and {Greenhill}, L.~J. and {Hall}, P.~J. and {Hazelton}, B.~J. and {Herne}, D. and {Hewitt}, J.~N. and {Johnston-Hollitt}, M. and {Kaplan}, D.~L. and {Kasper}, J.~C. and {Kincaid}, B.~B. and {Koenig}, R. and {Kratzenberg}, E. and {Lynch}, M.~J. and {Mckinley}, B. and {Mcwhirter}, S.~R. and {Morgan}, E. and {Oberoi}, D. and {Pathikulangara}, J. and {Prabu}, T. and {Remillard}, R.~A. and {Rogers}, A.~E.~E. and {Roshi}, A. and {Salah}, J.~E. and {Sault}, R.~J. and {Udaya-Shankar}, N. and {Schlagenhaufer}, F. and {Srivani}, K.~S. and {Stevens}, J. and {Subrahmanyan}, R. and {Waterson}, M. and {Webster}, R.~L. and {Whitney}, A.~R. and {Williams}, A. and {Williams}, C.~L. and {Wyithe}, J.~S.~B.},
        title = "{The Murchison Widefield Array: The Square Kilometre Array Precursor at Low Radio Frequencies}",
      journal = {\pasa},
         year = 2013,
        month = jan,
       volume = {30},
          eid = {e007},
        pages = {e007},
          doi = {10.1017/pasa.2012.007},
archivePrefix = {arXiv},
       eprint = {1206.6945},
 primaryClass = {astro-ph.IM},
       adsurl = {https://ui.adsabs.harvard.edu/abs/2013PASA...30....7T}
}

@ARTICLE{Teyssier2002AA,
       author = {{Teyssier}, R.},
        title = "{Cosmological hydrodynamics with adaptive mesh refinement. A new high resolution code called RAMSES}",
      journal = {\aap},
         year = 2002,
        month = apr,
       volume = {385},
        pages = {337-364},
          doi = {10.1051/0004-6361:20011817},
archivePrefix = {arXiv},
       eprint = {astro-ph/0111367},
 primaryClass = {astro-ph},
       adsurl = {https://ui.adsabs.harvard.edu/abs/2002A&A...385..337T}
}
